# Evidence for longitudinal variability of ethane ice on the surface of Pluto

B.J. Holler[a,b], L.A. Young[b,c], W.M. Grundy[b,d], C.B. Olkin[b,c], J.C. Cook[c]

a. Laboratory for Atmospheric and Space Physics, University of Colorado at Boulder, 1234 Innovation Dr., Boulder, CO 80303.
b. Remote observer at the Infrared Telescope Facility, which is operated by the University of Hawaii under Cooperative Agreement #NNX-08AE38A with the National Aeronautics and Space Administration, Science Mission Directorate, Planetary Astronomy Program.
c. Southwest Research Institute, 1050 Walnut St. #300, Boulder, CO 80302.
d. Lowell Observatory, 1400 W. Mars Hill Rd., Flagstaff, AZ 86001.

| | |
|---|---|
| Corresponding author: | Bryan Holler |
| Mailing address: | Laboratory for Atmospheric and Space Physics, University of Colorado at Boulder, 1234 Innovation Dr., Boulder, CO 80303. |
| E-mail: | bryan.holler@colorado.edu |
| Voice: | 302-383-8094 |
| Fax: | N/A |
| | |
| Running head: | Ethane Variability on Pluto |
| Manuscript pages: | 22 |
| Figures: | 7 |
| Tables: | 2 |
| | |
| Icarus keywords: | Pluto, surface; Ices, IR spectroscopy; Photochemistry; Pluto; Spectroscopy |

## *Abstract*

We present the results of an investigation using near-infrared spectra of Pluto taken on 72 separate nights using SpeX/IRTF. These data were obtained between 2001 and 2013 at various sub-observer longitudes. The aim of this work was to confirm the presence of ethane ice and to determine any longitudinal trends on the surface of Pluto. We computed models of the continuum near the 2.405 μm band using Hapke theory and calculated an equivalent width of the ethane absorption feature for six evenly-spaced longitude bins and a grand average spectrum. The 2.405 μm band on Pluto was detected at the 7.5-σ level from the grand average spectrum. Additionally, the band was found to vary longitudinally with the highest absorption occurring in the $N_2$-rich region and the lowest absorption occurring in the visibly dark region. The longitudinal variability of $^{12}CO$ does not match that of the 2.405 μm band, suggesting a minimal contribution to the band by $^{13}CO$. We argue for ethane production in the atmosphere and present a theory of volatile transport to explain the observed longitudinal trend.

## *1. Introduction*

Almost 85 years have passed since the discovery of Pluto, yet its surface characteristics are still not fully understood. The primary surface ice components are $N_2$, $CH_4$, and CO (Cruikshank et al. 1997). The presence of these volatile ices is elegantly explained by comparing Pluto's surface temperature (~40 K; Tryka et al. 1994; Lellouch et al. 2000, 2011) and diameter (~2368 km; Lellouch et al. 2014) to the volatile loss curves of $N_2$, $CH_4$, and CO (Schaller & Brown 2007). Pluto is sufficiently large and cold to retain these species over the age of the solar system and is able to support an atmosphere of $N_2$, $CH_4$, and CO (e.g.: Elliot et al. 2007). However, these volatile ices vary both with Pluto longitude and time (Grundy et al. 2013, hereafter referred to as G13; Grundy et al. 2014, hereafter referred to as G14). These variations are most likely due to changes in illumination across Pluto's surface as it orbits the Sun, allowing for sublimation of volatiles and subsequent transport, or to changes in viewing geometry.

In addition to Pluto's changing axial orientation with respect to the Sun and Earth, chemical processes are altering the composition of the surface. Extreme-UV photons and cosmic rays interact with molecules in the atmosphere, on the surface, and in some cases can penetrate deeper into the ice. In particular, $CH_4$ molecules may undergo photolysis or radiolysis to be converted into other hydrocarbon products such as acetylene ($C_2H_2$), ethylene ($C_2H_4$), ethane ($C_2H_6$), and propane ($C_3H_8$) (Lara et al. 1997; Krasnopolsky & Cruikshank 1999; Moore and Hudson 2003). From Figure 32 in Fray and Schmitt (2009), the sublimation pressures (at 40 K) of $N_2$ (~100 μbar), CO (~10 μbar), and $CH_4$ (~0.01 μbar) are much higher than those of acetylene, ethylene, and ethane (<<0.001 μbar). Non-methane hydrocarbon species shall henceforth be referred to as non-volatiles since their sublimation pressures are negligible at 40 K.

A simple calculation of the flux of Lyman-α photons reaching the surface of Pluto can answer the question of where photochemistry takes place: On the surface or in the atmosphere? The flux of photons ($F=F_0 e^{-\sigma N}$) hitting Pluto's surface depends on the photon flux at Pluto's orbital distance of 30 AU ($F_0=3 \times 10^8$ cm$^{-2}$ s$^{-1}$; Madey et al. 2002), the UV cross section of $CH_4$ at 120 nm ($\sigma=1.8 \times 10^{-17}$ cm$^2$; Chen and Wu 2004), and the column density of $CH_4$ ($N=1.75 \times 10^{19}$ cm$^{-2}$; Lellouch et al. 2009). This calculation yields a flux of Lyman-α photons on the order of $10^{-129}$ cm$^{-2}$ s$^{-1}$, a number that is effectively zero. If Pluto's atmosphere collapses, this calculation is no longer valid. Thus we assume photochemical products such as ethane are mainly formed in the atmosphere (Lara et al. 1997; Krasnopolsky & Cruikshank 1999) and the descent time is long enough that horizontal winds may transport the products a great distance away from the region of origin before precipitating onto the surface (Mark Bullock, private communication). This will result in a uniform surface distribution of photochemical products regardless of whether the atmospheric $CH_4$ is uniform (Lellouch et al. 2014) or not (Cook et al. 2013). These non-volatile ices may subsequently be covered over time by deposition of volatiles onto the surface.

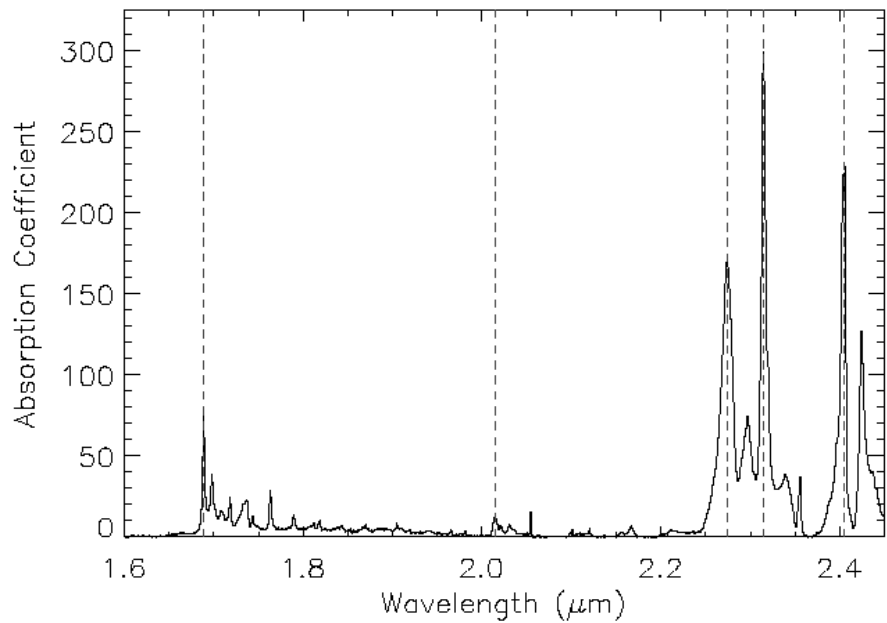

**Figure 1.** Absorption coefficients for pure ethane ($C_2H_6$) at 21 K (adapted from Quirico and Schmitt 1997a). Dashed gray lines mark the positions of ethane lines of interest in the range 1.6-2.45 μm. From left to right: 1.689, 2.015, 2.274, 2.314, and 2.405 μm. The spectral range of IRTF/SpeX is 0.8-2.43 μm.

Sasaki et al. (2005) performed a search for acetylene, ethylene, ethane, and propane ices on Pluto in the *L* band (3.0-4.0 μm) but the results were inconclusive. DeMeo et al. (2010) identified weak ethane absorption bands at 2.274, 2.405, 2.457, and 2.461 μm in the *K* band and constrained pure ethane to <10%. The 2.405 μm band coincides with an absorption band of $^{13}$CO, an isotopologue of the more abundant $^{12}$CO (Cruikshank et al. 2006). DeMeo et al. (2010) argue that the 2.405 μm band is too deep compared to the 2.457 and 2.461 μm bands, and therefore $^{13}$CO must contribute to the depth of the 2.405 μm band. However, individual bands of a species may not reach maximum depth at the same longitude (as is the case for $CH_4$ on Pluto and Triton from G13 and Grundy et al (2010), respectively). Conversely, Cruikshank et al. (2006) argue, based on currently unpublished CO laboratory data, that the contribution of $^{13}$CO is negligible and that the 2.405 μm feature is almost entirely due to ethane absorption. This issue will be addressed more thoroughly in the Discussion section. In the same manner as DeMeo et al. (2010), Merlin et al. (2010) find less pure ethane ice on Pluto's surface (5%) in favor of more heavily radiation-processed tholins (20%). They also present an ethane life cycle theory where the surface of Pluto is effectively shielded from radiation and cosmic rays by the atmosphere during perihelion and covered in $N_2$ ice during aphelion. They indicate a preference for ethane creation either on methane-rich surface ice patches during aphelion or within the atmosphere during perihelion. See Figure 1 for ethane bands relevant to this work and Table 1 in Hudson et al. (2009) for a full description of ethane absorption bands seen in the infrared.

## *2. Observations*

The combined Pluto/Charon spectra analyzed in this investigation were obtained on 72 nights from 2001 to 2013 using the SpeX infrared spectrograph on the 3-meter Infrared Telescope Facility (IRTF) (Rayner et al. 1998, 2003). The reader is referred to Table 1 in G13 for observational circumstances of the first 65 nights (2001-2012), and Table 1 in G14 for the observational circumstances of the later seven nights (2013). The observed wavelength range covered 0.8 to 2.43 μm using slit widths of 0.3" (λ/Δλ~1600-1900) and 0.5" (λ/Δλ~1200); the slit is 15" in length. Pluto itself subtends 0.1" while the maximum separation between Pluto and Charon is 1", too small for SpeX to routinely spatially resolve the two bodies. Charon accounts for 20.8% of the total reflecting area in the Pluto system. However, Charon contributes less than 20.8% of the light in a combined Pluto/Charon spectrum since Charon's albedo is wavelength-dependent and generally lower than Pluto's between 0.8 and 2.43 μm (Douté et al. 1999). Spectra were obtained with the slit rotation parallel to the imaginary line connecting Pluto and Charon so that the fraction of light from Charon was independent of slit width, seeing, or guiding accuracy. This eliminated the need to quantify Charon's contribution for each individual spectrum taken throughout a given night. We included Charon's contribution when performing spectral modeling. Pluto's minor satellites Nix, Styx, Kerberos, and Hydra are so small as to be negligible in this analysis (Weaver et al. 2006; Showalter et al. 2011, 2012). For a more thorough description of our observing process, the reader is again referred to G13 and G14.

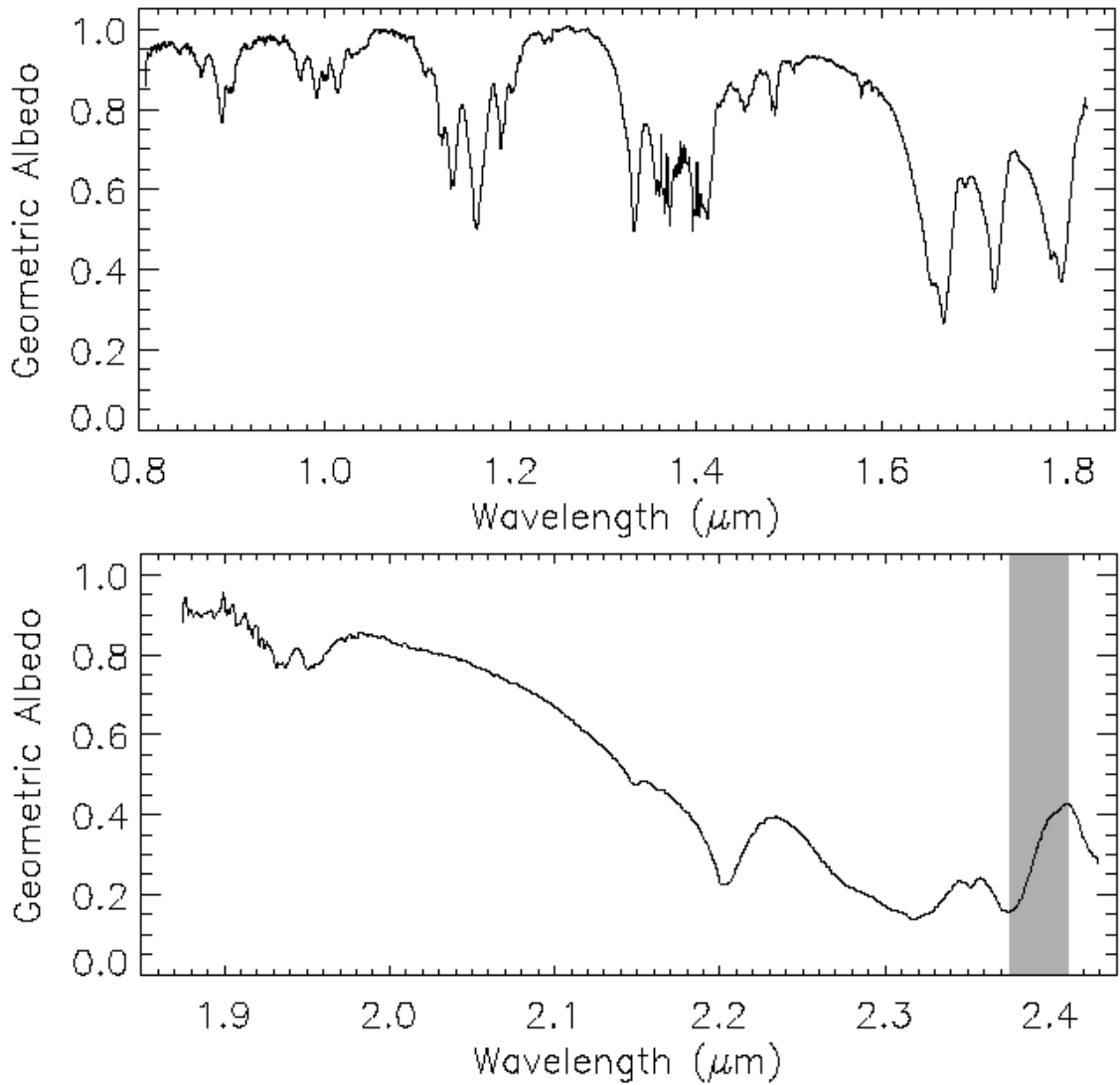

**Figure 2.** Combined Pluto/Charon grand average spectrum of the 72 individual spectra covering 0.8-1.85 μm (top panel) and 1.85-2.43 μm (bottom panel). Telluric absorption is present near 1.85 μm, therefore this region has been removed from the spectrum. The gray shaded region denotes the analysis region used in this investigation (described in the Analysis section). The signal-to-noise ratio of the grand average spectrum is 155.

## 3. Analysis

The raw spectra were reduced as described in G13. The reduced spectra analyzed in this investigation can be found as supplementary material accompanying this paper and at *http://www2.lowell.edu/users/grundy/abstracts/2014.IRTF-Pluto.html*. A weighted average was performed on the albedo values within each wavelength interval; each value was weighted according to its uncertainty, with more accurate measurements given larger weighting factors. The resulting grand average spectrum, calculated from 72 individual spectra and covering 0.8-2.43 µm, has a spectral resolution of ~1100 and a signal-to-noise ratio (SNR) of 155 and is seen in Figure 2. The SNR was calculated by fitting a portion of the spectrum (2.38-2.40 µm) to a cubic polynomial and evaluating the scatter of the data points with respect to the fit. This region was chosen to calculate the SNR because it not only was a good fit to a cubic but also comprised a large portion of the region included in the later analysis. The 72 spectra were then sorted into six longitude bins covering the 60° intervals described in Table 1. The spectra in each longitude bin were averaged in the same manner as the grand average. The bin ranges were chosen based on the spectral characteristics described in Figure 3. Bins 1 and 2 roughly match the section observed to be dark in Pluto's visible light curve (Buie et al. 2010a,b); this region is most likely dominated by low-albedo tholins, but this has yet to be confirmed. Bins 3 and 4 roughly correspond to a region of Pluto dominated by $N_2$ ice with a peak in absorption of CO, two species found to be spatially concurrent on Pluto (G13); the visible light curve also peaks in this region (Buie et al. 2010a,b). Bins 5 and 6 cover the third of Pluto dominated by $CH_4$ ice.

**Table 1**
Longitude Bins

| Bin # | Terrain Description | Longitude Range[1] | # of Spectra | SNR | W ($10^{-4}$ µm) |
|---|---|---|---|---|---|
| GA | Grand Average | 2.8°-355.4° | 72 | 155 | 1.57±0.21 |
| 1 | Dark 1 | 0°-60° | 9 | 55 | 2.14±0.57 |
| 2 | Dark 2 | 60°-120° | 13 | 108 | 0.72±0.38 |
| 3 | $N_2$-dominated 1 | 120°-180° | 13 | 51 | 2.18±0.52 |
| 4 | $N_2$-dominated 2 | 180°-240° | 15 | 68 | 1.71±0.42 |
| 5 | $CH_4$-dominated 1 | 240°-300° | 10 | 66 | 1.27±0.63 |
| 6 | $CH_4$-dominated 2 | 300°-360° | 12 | 91 | 1.11±0.50 |

[1]These longitudes correspond to the right-hand-rule coordinate system where 0° longitude is the sub-Charon point, north is aligned parallel to Pluto's spin axis, and sunrise is to the east.

The goal of this investigation was to determine how ethane abundance varies as a function of longitude across Pluto's surface. Our analysis focused on the 2.405 μm band. Other potential ethane bands within the range of the data (0.8-2.43 μm) at 1.689, 2.015, 2.274, and 2.314 μm suffered from confusion by strong $CH_4$ absorption bands or telluric absorption. More bands exist at 2.457 and 2.461 μm but fell outside of the data range. We made use of code based on Hapke theory (e.g.: Hapke 2012) to construct synthetic spectra that modeled the region near the 2.405 μm band. For convenience, these models will be referred to henceforth as ethane-absent continuum models. They are not spectrally flat, but instead are models assuming no ethane present on Pluto's surface. This was a more appropriate method than simply fitting the continuum to a polynomial because the surface ice components and their characteristics were taken into account. Additionally, the polynomial fits did not appear adequate even to the eye. The equivalent width of an absorption band provides a reasonable stand-in for the abundance of a substance present so we calculated the equivalent width of the 2.405 μm band for each longitude bin and the grand average using the normalized residual.

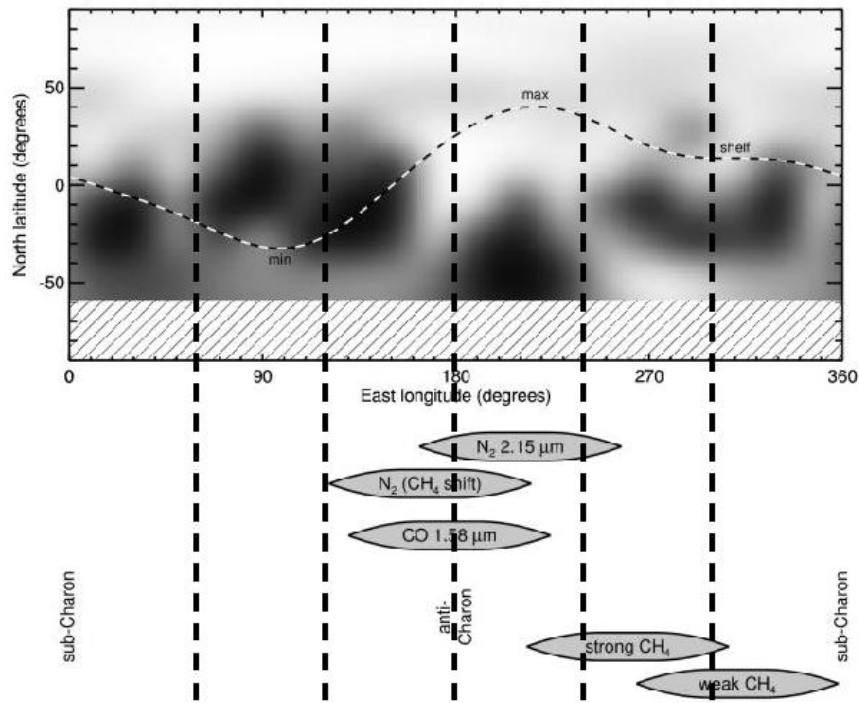

**Figure 3.** Pluto albedo map and visible light curve (Buie et al. 2010a,b) with approximate longitude regions of peak absorption marked for various species by gray "battleships". This plot is adapted from Figure 25 in GR13 with dashed lines added to show the regions covered by the six 60° longitude bins. The hash marks denote the region of Pluto not currently observable at the time of the Buie et al. data acquisition (2002-2003). The region from 0°-90° is dark and does not show a peak absorption of any particular species. Strong and weak refer to strength of the $CH_4$ bands. The strong $CH_4$ bands have a peak absorption offset spatially from the weak $CH_4$ bands. See G13 for descriptions of the strong and weak $CH_4$ bands.

Our analysis spanned the 2.376-2.410 μm region (gray shaded region in Figure 2); more points were present at shorter and longer wavelengths but were never adequately fit by the synthetic spectra, resulting in a visibly worse model for the continuum in the region of interest near the 2.405 μm band. Decreasing the number of points also decreased the computation time significantly. The points within the band (2.399-2.409 μm; ~10 nm in width) were excluded from the ethane-absent continuum models so that discrepancies between the model and data would not influence the $\chi^2$ value and the behavior of the code. Failing to remove these points usually resulted in a model with parameters for the other materials altered in an attempt to fit the shape of the band.

We followed the procedure used in Cook et al. (2007) using modeling techniques described by Roush (1994) for computing the ethane-absent continuum models. The code used a simplex method (e.g.: Press et al. 2007) to search the multi-dimensional parameter space for the combination of mass fractions and grain sizes that minimized $\chi^2$. Individual materials were placed in intimate mixtures (also called "salt-and-pepper" mixtures) making up separate spatial components. We considered four spatial components in our ethane-absent continuum models: pure $CH_4$, Triton tholin, an $N_2$/$N_2$:CO/shifted $CH_4$ intimate mixture, and an $H_2O$/amorphous carbon intimate mixture (to model Charon). The areal coverage of the Charon spatial component was set to a constant value of 20.8% throughout this investigation based on the ratio of Charon's reflecting area to Pluto's. The areal coverage of the other three spatial components was allowed to vary.

**Table 2**
Optical Constants

| Material | Temperature (K) | Reference |
|---|---|---|
| $CH_4$ | 40 | Grundy et al. (2002) |
| Triton tholin | N/A | Khare et al. (1994) |
| $N_2$ | 40 | Grundy et al. (1993) |
| $N_2$:CO | 36.5 | Quirico and Schmitt (1997b) |
| Amorphous carbon | N/A | Rouleau and Martin (1991) |
| $H_2O$ | 50 | Grundy and Schmitt (1998) |

The sources of the optical constants of the materials considered in this investigation can be found in Table 2. Optical constants are temperature-dependent and affect the absorptive and reflective properties of a given material. The temperature of the $N_2$ and $CH_4$ ices could be chosen manually, resulting in a re-calculation of the optical constants; these are the temperature values shown in Table 2. The $N_2$ and $CH_4$ temperatures (40 K) were chosen based on estimates of Pluto's surface temperature (~40 K; Tryka et al. 1994; Lellouch et al. 2000, 2011). Additionally, $CH_4$ could be given an arbitrary spectral blueshift to account for being dissolved in solution with $N_2$. By manually choosing a blueshift, optical constants for pure $CH_4$ were used instead of those for an $N_2$:$CH_4$ solution. A constant blueshift of 6 nm was chosen for the shifted $CH_4$ throughout this investigation for the purpose of consistency.

Because no $N_2$ or CO bands were present in the small region analyzed, the 2.15 μm $N_2$ and 2.35 μm CO bands were analyzed independently to determine the best-fit mass fractions and grain sizes. For $N_2$, the best-fit values were 99.17% and 103.6 mm, respectively. The best-fit values for CO were 0.1048% and 0.2585 mm, respectively. The errors on these values play no role from this point forward so are not reported here. Both DeMeo et al. (2010) and Merlin et al. (2010) chose an $N_2$ mass fraction of 99.54% and a CO mass fraction of 0.1%. DeMeo et al. (2010) chose an $N_2$ and CO grain size of 95 mm. The $N_2$ and CO grain sizes are identical since they do not consider an intimate mixture but instead have CO dissolved in solution with $N_2$. This investigation also used an $N_2$:CO solution, and the different grain sizes are interpreted as due to the separation of CO reflecting centers within the larger $N_2$ matrix. Similarly, Merlin et al. (2010) use an $N_2$ and CO solution and take the grain size to be 20 cm. This large grain size is most likely due to individual particles of sub-millimeter size behaving as an aggregate (Grundy and Buie 2001). The mass fractions and grain sizes of $N_2$ and CO were then set as constants throughout the analysis of the selected region. The mass fractions and grain sizes of the other materials remained free parameters.

Synthetic spectra created using Hapke theory are not unique: Degeneracies exist between the mass fraction and grain size of a particular material and those of the other materials. No special significance should be given to the values output by the Hapke code beyond the fact that the values result in a reasonable fit to the data. Areal coverage for the spatial components, mass fractions, and grain sizes were computed for the materials described above but have limited scientific value since they are based only on an analysis of the region from 2.376-2.410 μm; they are not quoted here. We do not perform a full analysis of the spectra from 0.8-2.43 μm as it is beyond the scope of this paper.

After obtaining the ethane-absent continuum model for a particular longitude bin, the residuals (model minus data) within the region of the ethane band were calculated. The residuals were then normalized by the ethane-absent continuum for consistency and easier comparison of results with G13 and G14. The equivalent width of the 2.405 μm band and associated errors were computed by performing a simple Riemann sum over the extent of the band (2.399-2.409 μm): W=$\Sigma \Delta_i R_i$, where $\Delta_i$ is the wavelength separation between data points $i$-$1$ and $i$ and $R_i$ is the normalized residual at data point $i$.

## *4. Results*

We report a 7.5-σ detection of the 2.405 μm band from modeling the continuum of the grand average spectrum. The equivalent width of the band was calculated to be (1.57±0.21) x $10^{-4}$ μm. The dashed line in Figure 4 denotes the grand average equivalent width value with the gray shaded regions representing the uncertainty. These SpeX data suggest the presence of ethane ice on the surface of Pluto.

Equivalent widths for the 2.405 μm band from the six longitude bins can be found in Table 1 and are presented graphically in Figure 4. Nominally, peak absorption occurs between 120°-180°, but due to the uncertainties on the equivalent width calculations, the peak may actually occur in either the 0°-60° (Dark 1) or 180°-240° ($N_2$-dominated 2) bins instead. The minimum could occur in either the 60°-120° (Dark 2), 240°-300° ($CH_4$-dominated 1), or 300°-360° ($CH_4$-dominated 2) bins. Pluto's visible light curve is at a minimum between 60° and 120° (Buie et al. 2010 a,b), a potentially tholin-dominated region that almost exactly coincides with the full extent of the darkest area of Pluto as seen in Figure 3. Absorption is not clearly evident in the bin 2 spectrum (Figure 5, Figure 6) and the equivalent width value is 1.9-σ above zero; a definitive statement about absorption in this region cannot be confidently made based on the statistics. Additionally, the difference between the equivalent width in this region and those of the adjacent regions is statistically significant. In terms of equivalent width of the 2.405 μm band, the 60°-120° region is unique on this half of Pluto and appears to have more in common with the $CH_4$-dominated regions (240°-300° and 300°-360°). The ethane-absent continuum models for all six longitude bins and the grand average plotted against the data are found in Figure 5 with normalized residuals found in Figure 6.

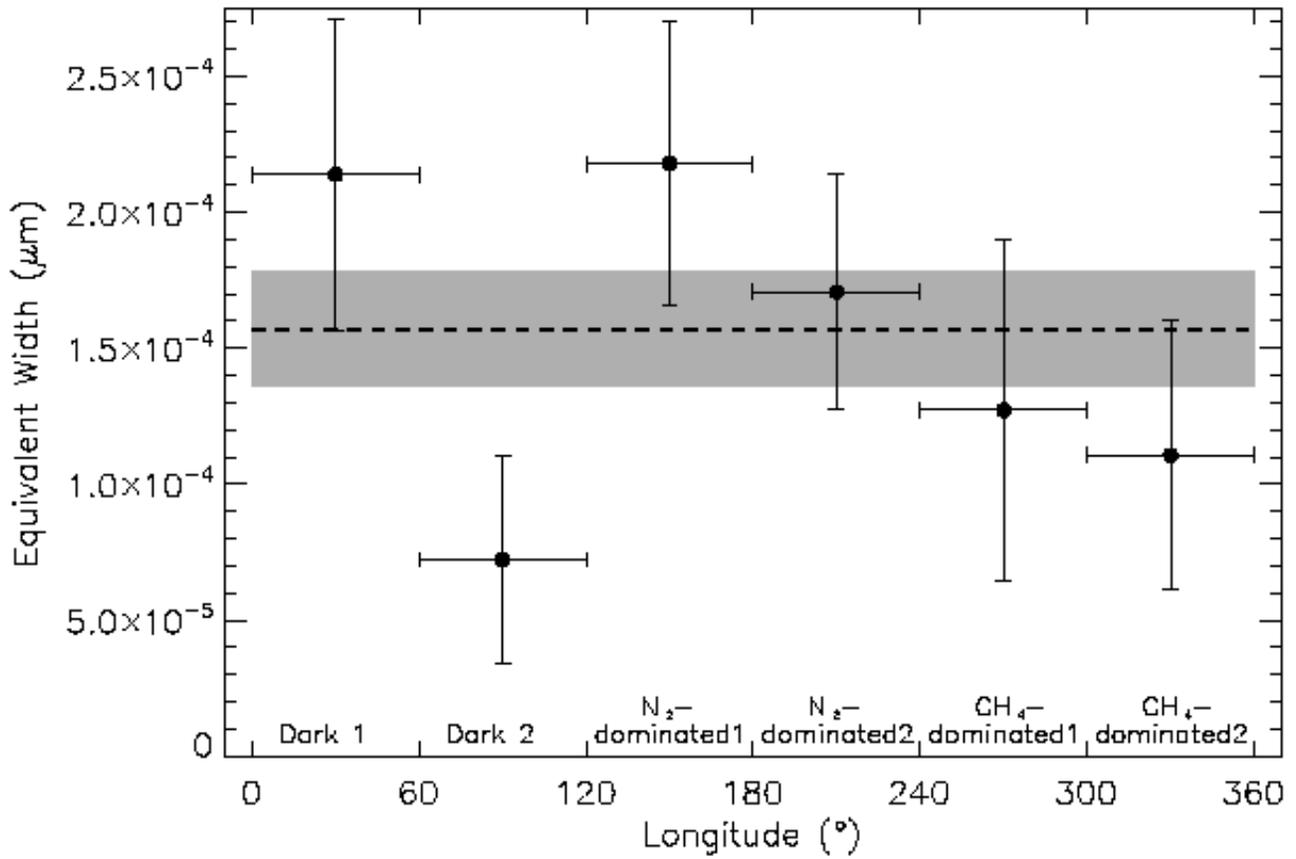

**Figure 4.** Equivalent width for the 2.405 μm band (attributed to ethane) in each of the 60° longitude bins. Vertical error bars are 1-σ errors computed on the equivalent width. Horizontal bars mark the longitude range covered by each bin. The black dashed line is the value of the equivalent width from the grand average and the gray area represents the error on that calculation.

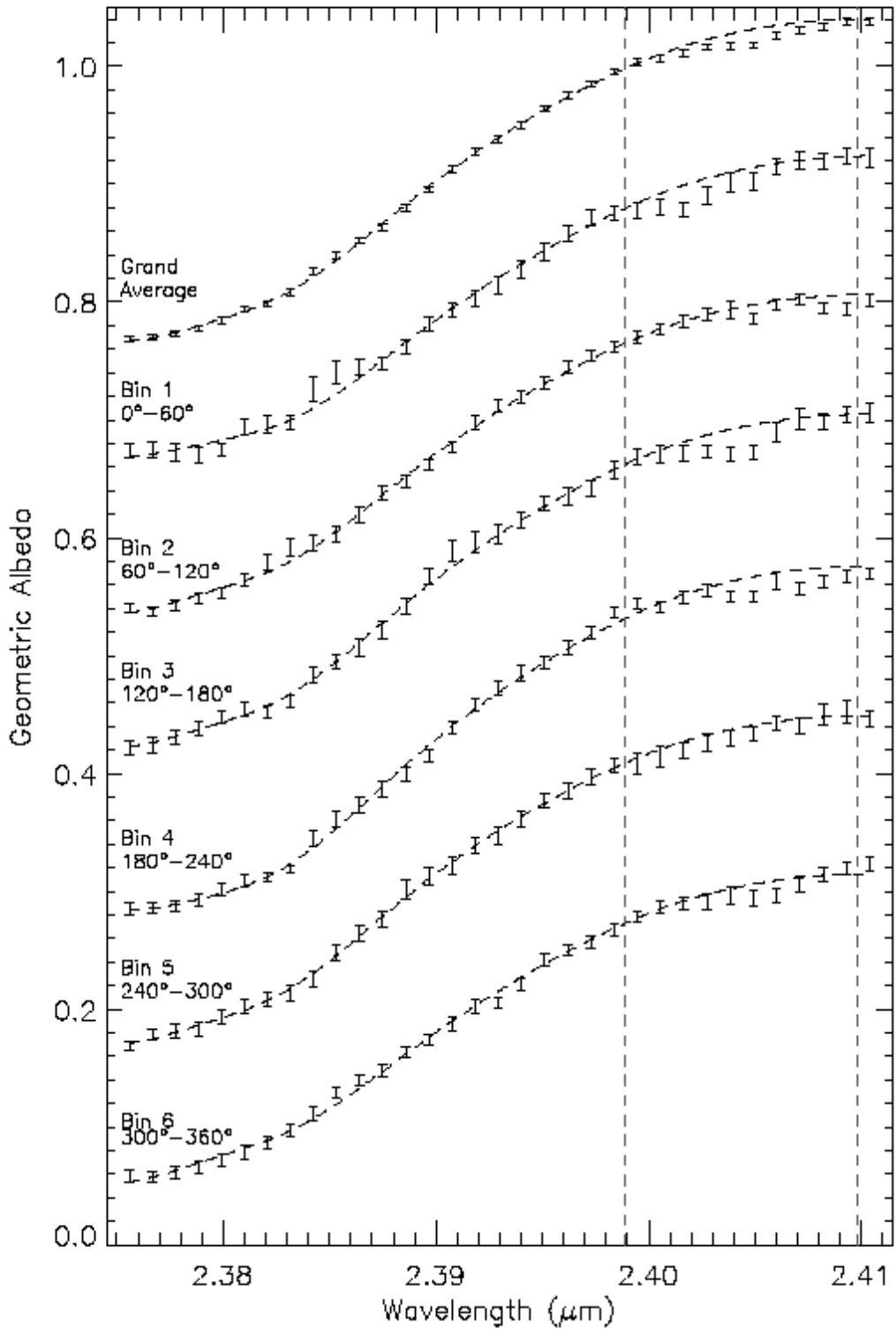

**Figure 5.** Stacked spectra of the grand and longitude bin averages from 2.376 to 2.410 μm. All spectra are offset from zero for clarity and do not represent the true geometric albedo values. Ethane-absent models are represented by the dashed lines and are unique for each longitude bin. Labels on the left-hand side refer to the curve directly below. Vertical dashed lines denote the extent of the 2.405 μm band (2.399-2.409 μm).

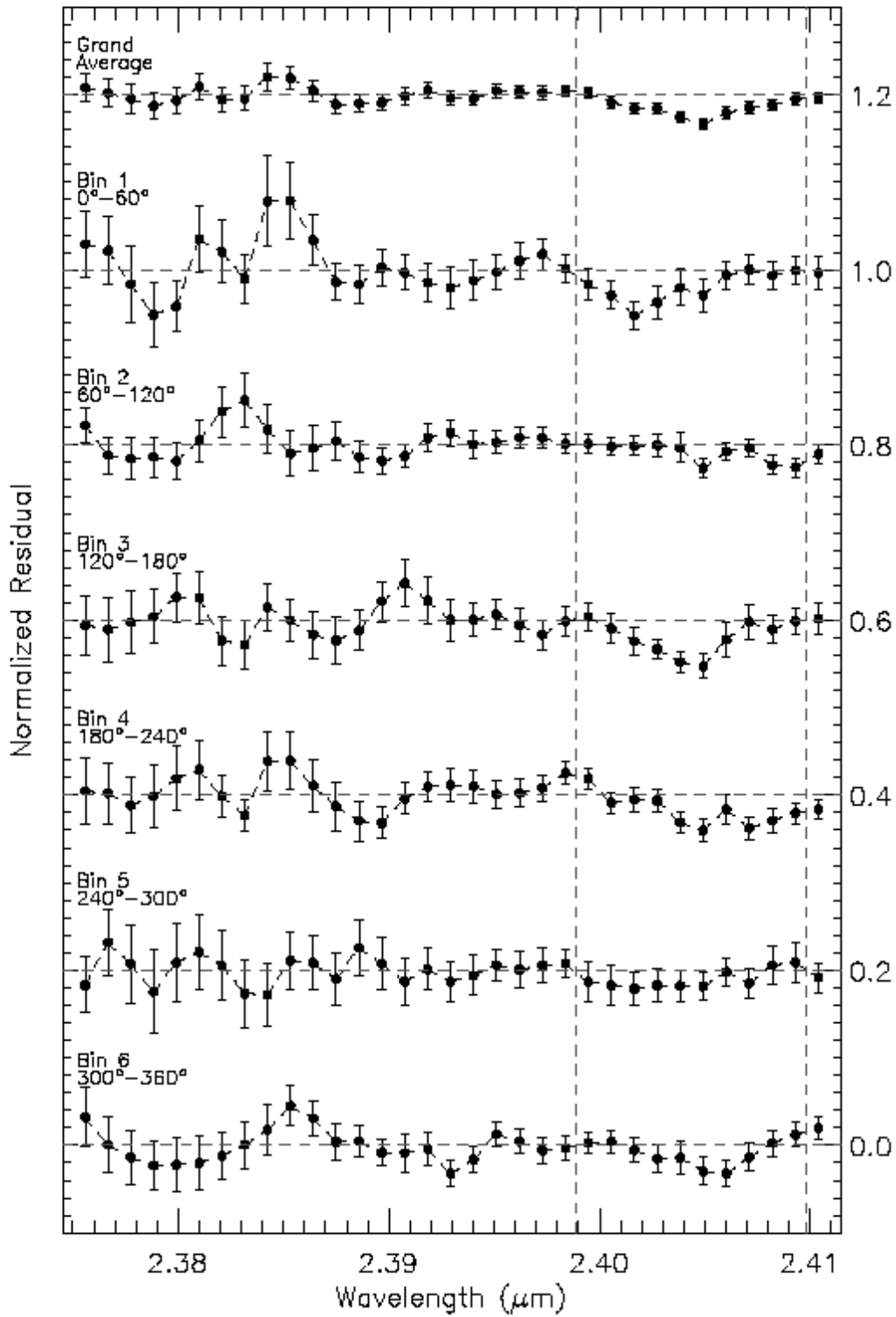

**Figure 6.** Stacked normalized residuals of the grand and longitude bin averages from 2.376 to 2.410 μm. Horizontal dashed lines show the zero value for each bin. Labels on the left-hand side refer to the curve directly below. Vertical dashed lines denote the extent of the 2.405 μm band (2.399-2.409 μm). Bins 3 (4.2-σ) and 4 (4.1-σ) show a definite detection of the band, while Bins 2 (1.9-σ) and 5 (2.0-σ) show a less certain detection of the band.

## 5. Discussion

As discussed briefly in the Introduction, $^{13}$CO and ethane both have absorption features at 2.405 μm, the band of interest in this investigation. Disentangling the contributions of these two species presents a problem. No other ethane bands were identified in our spectra to corroborate a detection of ethane ice on the surface of Pluto, either due to confusion by $CH_4$ absorption or because the bands are found beyond the spectral range of the SpeX instrument. To solve this problem, we chose to look into the longitudinal distribution of $^{12}$CO from the equivalent widths of the 1.58 μm $^{12}$CO absorption feature. The 2.35 μm feature is not useful due to its location between two strong $CH_4$ bands. Without a way to confidently quantify the fractionation of $^{13}$CO and $^{12}$CO ices on the surface of Pluto, we assumed that these two isotopologues have identical spatial distributions. Determining the spatial distribution of $^{12}$CO therefore indirectly provides a spatial distribution for $^{13}$CO.

Figure 3 from G13 shows the longitudinal distribution of $^{12}$CO from calculations of the equivalent widths of the 1.58 μm feature in nightly spectra. The distribution could be described as "triangular" with a peak near 180° longitude. Compare this to the pattern of equivalent widths for the 2.405 μm band in Figure 4 of this work. The two distributions do not match, so we conclude that the contribution of $^{13}$CO to the observed depth of the 2.405 μm band is minimal. Due to the scarcity of $^{13}$CO laboratory optical constants and absorption coefficients, we are unable to calculate a ratio of ethane to $^{13}$CO for the 2.405 μm band. We are also unable to evaluate the accuracy of our assumptions about the identical spatial distributions of the CO isotopologues. That being said, we will move forward with the interpretation of our results assuming that the 2.405 μm band is due almost entirely to ethane absorption.

The 2.405 μm band is not likely due to $CH_4$ absorption either. New absorption coefficients for $CH_4:N_2$ and $N_2:CH_4$ (Protopapa et al. 2013) do not show a band at 2.405 μm. However, $CH_4$ absorption occurs at shorter and longer wavelengths on either side of the 2.405 μm feature. The $CH_4$ absorption vs. longitude distribution is not anti-correlated to that of ethane (compare Figure 4 of this work to Figure 4 in G13). Anti-correlation would imply that the variations in the 2.405 μm equivalent width are artificial and due to variation in nearby $CH_4$ band depths with longitude. The two distributions are not anti-correlated, suggesting that the variation seen in the 2.405 μm band are due to variations in ethane absorption across Pluto's surface.

Ethane is a non-volatile species at outer solar system temperatures, so any ethane created in the atmosphere should remain on Pluto until converted to other hydrocarbon compounds. The longitudinal variations detected in this investigation suggests that volatile transport plays a key role in making Pluto's surface a dynamic milieu. Compare the nominal peak of ethane absorption (120°-180°) to the peak of the visible light curve (180°-240°) (Buie et al. 2010a,b) in Figure 7. The maximum ethane absorption is shifted one longitude bin to the west of the light curve peak, whereas the other bins approximately follow the same pattern as the light curve. Note also that the data used to construct the albedo map and the visible light curve of Pluto were obtained between 2002 and 2003, and that a majority of our SpeX data were obtained after this period. As described in the Introduction, we would expect to see a uniform distribution of ethane on the surface if formed in the atmosphere. However, this does not describe the observation of a non-uniform longitudinal distribution with a peak in the $N_2$-dominated 1 (120°-180°) region. This does not mean that ethane is not formed in the atmosphere, but instead that we must develop a more sophisticated explanation.

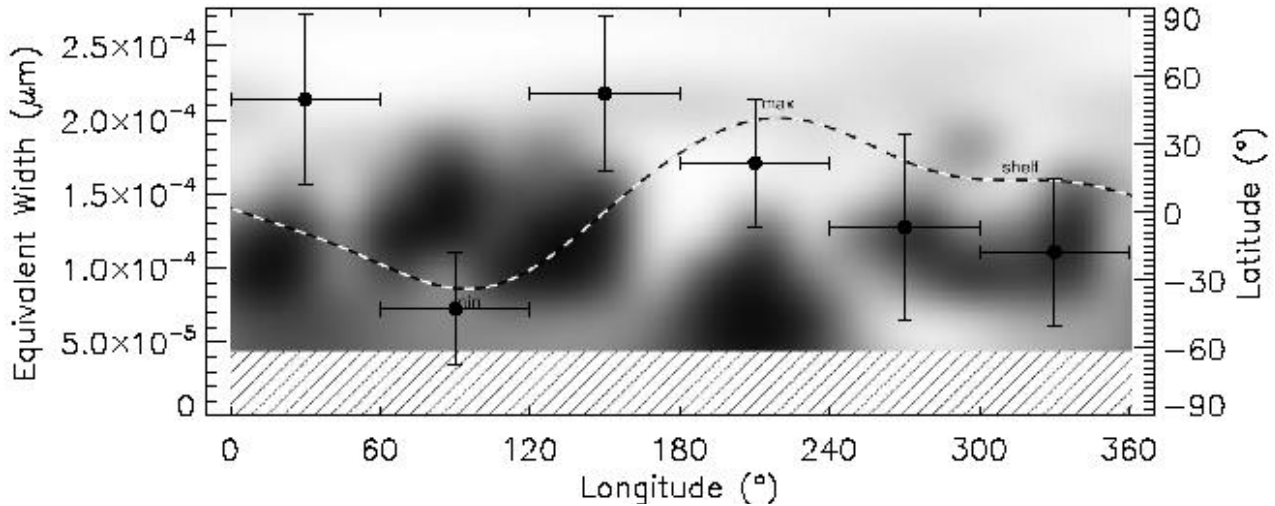

**Figure 7.** Equivalent width values from Table 1/Figure 4 super-imposed on the Pluto albedo map and visible light curve of Buie et al. (2010a,b). The visible light curve albedo is arbitrarily scaled. Refer to Figure 3 for more description of the albedo map and Figure 4 for more description of the equivalent widths.

We suggest that, as Pluto's North Pole continues to rotate into view following perihelion in 1989, the northern latitudes are receiving an increasing amount of solar insolation. These northern latitudes are home to the majority of the high-albedo volatile ices on Pluto, as seen in the global albedo map (Buie et al. 2010a,b). As these latitudes receive more energy, the deposited volatiles (predominantly $N_2$ between 120° and 240° longitude) sublimate and either reside in the atmosphere or are transported and re-deposited in other regions. The sublimation rate has increased faster than the deposition rate as revealed by Pluto's monotonically increasing atmospheric pressure between 1989 and 2013 (Olkin et al. 2014). As the volatile ices sublimate from the northern latitudes, the underlying non-volatile ethane is revealed. In this scenario, the shift between the absorption and light curve peaks is due to the original amount of volatiles present in these regions. We expect the brighter areas to be correlated with more volatile ice content and to absorb less sunlight, therefore sublimation in this region proceeds more slowly. A greater quantity of volatiles should remain in the region marked by a peak in visible albedo. This means the region directly to the west may have undergone significant uncovering of the underlying ethane. Conversely, the regions dominated by $CH_4$ show less ethane absorption because $CH_4$ is less volatile than both $N_2$ and CO. What we are indirectly seeing is evidence for changes in Pluto's regional albedos and volatile distribution since the albedo map data were obtained.

The $CH_4$-dominated regions (240°-300° and 300°-360°) are not likely candidates as regions of peak ethane absorption. The difference between the equivalent widths of these bins and that of the $N_2$-dominated 1 bin (120°-180°) are at the 1.1-σ and 1.5-σ levels, respectively. This is further evidence that ethane forms primarily in Pluto's atmosphere instead of on the surface. Ethane formation in the surface ice is not impossible, but it should make a negligible contribution to the total ethane abundance due to atmospheric shielding, as seen from the calculation performed in the Introduction. Again, if Pluto's atmosphere collapses, the previous statement is no longer accurate and ethane formation in the surface ice would be the dominant production mechanism. Ethane formation will occur when $CH_4$ is present in the atmosphere and may be ongoing throughout a full Pluto orbit at different rates. Maximum ethane production in the atmosphere would be expected to peak post-perihelion at maximum atmospheric pressure. Pluto's atmospheric pressure continues to monotonically increase (Olkin et al. 2014) as it moves away from perihelion, so it is possible that the peak ethane formation period has not yet occurred.

The Ralph/LEISA instrument on New Horizons has the spectral range (1.25-2.50 μm) and spectral resolution (250) to resolve the 2.405 μm band on Pluto (Figure 6 in Young et al. 2008). Ralph/LEISA will be able to confirm longitudinal variability of ethane with spatial resolution upwards of 5-7 km pixel$^{-1}$ near closest approach to Pluto (Young et al. 2008). The Ralph/MVIC (four filters: blue (400-550 nm), red (540-700 nm), near-IR (780-975 nm), $CH_4$ filter (860-910 nm)) and LORRI (wavelength range of 350-850 nm) instruments onboard New Horizons will make available color and albedo information (Young et al. 2008; Cheng et al. 2008) useful for testing our theory of volatile transport on Pluto. The albedo data will be useful for constructing an updated global albedo map for comparison to that of Buie et al. (2010a,b). Color data from these instruments will suggest potential sublimation and deposition sites on Pluto (Moore et al. 1999; Howard and Moore 2008). Bluer regions might be considered sites of recent deposition, with redder regions potentially marking sites of recent sublimation as noticed on Triton from Voyager 2 data (Eluszkiewicz 1991). The Alice instrument will be useful for determining the height at which formation occurs by obtaining a vertical profile of gaseous ethane and $CH_4$ mixing ratios.

## *6. Conclusion*

Using near-infrared spectra of Pluto taken with the SpeX instrument at the IRTF over 72 nights, our investigation has made an independent confirmation of the 2.405 μm band at the 7.5-σ level, hinting heavily at the presence of ethane ice on Pluto's surface. We detect longitudinal variability of ethane by breaking the surface into six 60° longitude bins approximately matching previously-noticed surface ice distributions. The longitudinal variability observed for the equivalent width of the 2.405 μm band does not match that of $^{12}CO$ and $^{13}CO$, suggesting that the band is due entirely to ethane absorption. Peak ethane absorption nominally occurs in the $N_2$-dominated region between 120° and 180° longitude, with a nominal minimum in the Dark 2 region between 60° and 120° longitude. Decreased ethane absorption in the methane dominated regions (240°-300° and 300°-360°) suggests ethane formation in the atmosphere and a uniform distribution of ethane across Pluto's surface. The non-uniformity of the ethane distribution points to coverage of non-volatile ethane by volatile ices, with transport of these volatiles as seasons on Pluto change. We predict that New Horizons will find evidence for recent sublimation on Pluto between the longitudes of 120° and 180°.

Our IRTF/SpeX data set currently constitutes a long time baseline of Pluto data taken at many sub-observer longitudes. We very eagerly anticipate the flyby of the New Horizons spacecraft through the Pluto system in July 2015. New Horizons will provide a much-needed "ground truth" comparison for post-flyby data obtained through our IRTF/SpeX program and with other ground-based instruments.

## *Acknowledgements*

The authors would like to specially thank Bobby Bus for his illuminating questions and advice during the 2013 DPS poster session. Mark Bullock deserves a special thanks for adding an extra and necessary dimension to the evaluation of the results of this paper. We graciously thank the staff of the IRTF for their assistance over the past 13 years, especially W. Golisch, D. Griep, P. Sears, E. Volquardsen, J.T. Rayner, A.T. Tokunaga, and B. Cabreira. We wish to recognize and acknowledge the significant cultural role and reverence of the summit of Mauna Kea within the indigenous Hawaiian community and to express our appreciation for the opportunity to observe from this special mountain. Special thanks also to Fran Bagenal, Dale Cruikshank, Francesca DeMeo, Stephen Tegler, and Eliot Young for guidance and advice. Many thanks to the two reviewers for their perceptive comments that were used to significantly strengthen the arguments made in this paper. This work was funded by NASA PAST NNX13AG06G.